\documentstyle[11pt,epsfig]{article}
\date{}
\begin{document}
\title{{\bf Dilaton Cosmology, Noncommutativity and Generalized Uncertainty Principle}}
\author{Babak Vakili\thanks{%
email: b-vakili@sbu.ac.ir} \\
{\small Department of Physics, Shahid Beheshti University, Evin,
Tehran 19839, Iran}} \maketitle

\begin{abstract}
The effects of noncommutativity and of the existence of a minimal
length on the phase space of a dilatonic cosmological model are
investigated. The existence of a minimum length, results in the
Generalized Uncertainty Principle (GUP), which is a deformed
Heisenberg algebra between the minisuperspace variables and their
momenta operators. We extend these deformed commutating relations to
the corresponding deformed Poisson algebra. For an exponential
dilaton potential, the exact classical and quantum solutions in the
commutative and noncommutative cases, and some approximate
analytical solutions in the case of GUP, are presented and compared.
\vspace{5mm}\newline PACS numbers: 04.60.-m, 04.60.Ds,
04.60.Kz\vspace{.5cm}
\end{abstract}
\section{Introduction}
Since cosmology can test physics at energies that are much higher
than those which the experiments on Earth can achieve, it seems
natural that the effects of quantum gravity could be observed in
this context. Therefore, until a completely satisfactory theory
regarding cosmology can be afforded by string theory, the study of
the general properties of quantum gravity through cosmological
systems such as the universe seems reasonably promising and in
recent years many efforts have been made in cosmology from string
theory point of view \cite{1}-\cite{4}. In the pre-big bang
scenario, based on the string effective action \cite{5}, the birth
of the universe is described by a transition from the string
perturbative vacuum with weak coupling, low curvature and cold state
to the standard radiation dominated regime, passing through a high
curvature and strong coupling phase. This transition is made by the
kinetic energy term of the dilaton, an scalar field with which the
Einstein- Hilbert action of general relativity is augmented, see
\cite{6} for a more modern review of string dilaton cosmology. One
of the major features of the solutions of equations of motion in
string dilaton cosmology (see for example \cite{7} for some exact
solutions in dilaton cosmology) is the duality, so that if $a(t)$,
the scale factor, solves the equations of motion, $1/a(t)$ is also a
solution. This means that the whole universe behaves like a string,
i. e. has a minimal size of order of string scale and also a maximal
size of order of the inverse of string scale.

The existence of a minimal length is one of the most important
predictions of the theories which deal with quantum gravity
\cite{8}. From perturbative string theory point of view, such a
minimal length is due to the fact that the strings cannot probe
distances smaller than the string size. One of the interesting
features of the existence of a minimal length described above is
the modification it makes to the standard commutation relation
between position and momentum in usual quantum mechanics
{\cite{9,10}}, which are called Generalized Uncertainty Principle
(GUP). In one dimension the simplest form of such relations can be
written as
\begin{equation}\label{AA}\bigtriangleup p \bigtriangleup x\geq
\frac{\hbar}{2}\left(1+\beta (\bigtriangleup
p)^2+\gamma\right),\end{equation} where $\beta$ and $\gamma$ are
positive and independent of $\bigtriangleup x$ and $\bigtriangleup
p$, but may in general depend on the expectation values $<x>$ and
$<p>$. The usual Heisenberg commutation relation can be recovered
in the limit $\beta=\gamma=0$. As is clear from equation
(\ref{AA}), this equation implies a minimum position uncertainty
of $(\bigtriangleup x)_{min}=\hbar \sqrt{\beta}$, and hence
$\beta$ must be related to the Planck length. Now, it is possible
to realize equation (\ref{AA}) from the following commutation
relation between position and momentum operators
\begin{equation}\label{BB} \left[x,p\right]=i\hbar \left(1+\beta
p^2\right),\end{equation} where we take $\gamma=\beta <p>^2$. More
general cases of such commutation relations are studied in Refs.
\cite{11}.

 One of interesting features of GUP in more than one
dimension is that it implies naturally a noncommutative geometric
generalization of position space \cite{9}. Noncommutativity
between spacetime coordinates was first introduced by Snyder
\cite{12}, and in more recent times a great deal of interest has
been generated in this area of research \cite{13}-\cite{15}. This
interest has been gathering pace in recent years because of strong
motivations in the development of string and M-theories,
{\cite{16, 17}}. However, noncommutative theories may also be
justified in their own right because of the interesting
predictions they have made in particle physics, a few examples of
which are the IR/UV mixing and non-locality \cite{18}, Lorentz
violation \cite{19} and new physics at very short distance scales
\cite{19}-\cite{21}. Noncommutative versions of ordinary quantum
\cite{22} and classical mechanics {\cite{23, 24}} have also been
studied and shown to be equivalent to their commutative versions
if an external magnetic field is added to the Hamiltonian.

In cosmological systems, since the scale factors, matter fields and
their conjugate momenta play the role of dynamical variables of the
system, introduction of noncommutativity by adopting the approach
discussed above is particularly relevant. The resulting
noncommutative classical and quantum cosmology of such models have
been studied in different works \cite{25}. These and similar works
have opened a new window through which some of problems related to
cosmology can be looked at and, hopefully, resolved. For example, an
investigation of the cosmological constant problem can be found in
\cite{26}. In \cite{27} the same problem is carried over to the
Kaluza-Klein cosmology. The problem of compactification and
stabilization of the extra dimensions in multidimensional cosmology
may also be addressed using noncommutative ideas in \cite{28}.

In this paper we deal with noncommutativity and GUP in a dilaton
cosmological model with an exponential dilaton potential and to
facilitate solutions for the case under consideration, we choose a
suitable metric. Our approach to GUP is through its introduction in
phase space constructed by minisuperspace fields and their conjugate
momenta \cite{29}. In general GUP in its original form (see
\cite{9,10}) implies a noncommutative underlying geometry for space
time. But formulation of gravity in a noncommutative space time is
highly nonlinear and setting up cosmological models is not an easy
task. Here our aim is to study some aspects regarding the
application of the GUP framework in quantum cosmology, {\it i. e.}
in the context of a minisuperspace reduction of the dynamics. As is
well-known in the minisuperspace approach of quantum cosmology,
which is based on the canonical quantization procedure, one first
freezes a large number of degrees of freedom by imposition of
symmetries on the spacial part of the metric and then quantizes the
remaining ones. Therefore, in the absence of a full theory of
quantum gravity, quantum cosmology is a quantum mechanical toy model
with a finite degrees of freedom which is a simple arena to test
ideas and constructions which can be introduced in quantum general
relativity. In this respect, the GUP approach to quantum cosmology
appears to have physical grounds. In fact, one notes that a
deformation of the canonical Heisenberg algebra immediately leads to
a generalized uncertainty principle. In other words, the GUP scheme
relies on a modification of the canonical quantization prescriptions
and, in this respect, it can be reliably applied to any dynamical
system (see \cite{Mon} for a more clear explanation on the GUP in
the minisuperspace dynamics). Since our model has two degrees of
freedom, the scale factor $a$ and the dilaton $\phi$, with a change
of variables, we have a set of dynamical variables $(x,y)$, which
are suitable candidates for introducing noncommutativity and GUP in
the phase space of the problem at hand. We present exact solutions
of classical and quantum commutative and noncommutative cosmology.
Also in the case when the minisuperspace variables obey the GUP
commutating relations, we obtain approximate analytical solutions
for the corresponding classical and quantum cosmology. Finally, we
compare and contrast these solutions at both classical and quantum
levels.
\section{The model}
In $D=4$ dimension lowest order gravi-dilaton effective action, in
the string frame, can be written as \cite{30}
\begin{equation}\label{A}
{\cal S}=-\frac{1}{2 \lambda_s}\int d^4x
\sqrt{-g}e^{-\phi}\left({\cal R}+\partial_{\mu}\phi
\partial^{\mu}\phi+V(\phi)\right),\end{equation}where $\phi$ is
the dilaton field, $\lambda_s$ is the fundamental string length
$l_s$ parameter and $V(\phi)$ is the dilaton potential. In the
string frame our fundamental unit is the string length $l_s$, and
thus the Planck mass, which is the effective coefficient of the
Ricci scalar ${\cal R}$, varies with the dilaton. One can also write
the action in the Einstein frame, for which the fundamental unit is
the Planck length. Since the Planck length is more appropriate for
our purpose, we prefer to work in the Einstein frame. In \cite{4},
it is shown in details that action (\ref{A}) in the Einstein frame
takes the form
\begin{equation}\label{B}
{\cal S}=-\frac{M_4^2}{2}\int d^4x \sqrt{-g}\left({\cal
R}-\frac{1}{2}\partial_{\mu}\phi
\partial^{\mu}\phi-V(\phi)\right),\end{equation}where now all
quantities in the action are in the Einstein frame. We consider a
spatially flat FRW spacetime which, following \cite{31}, is
specified by the metric
\begin{equation}\label{C}
ds^2=-\frac{N^2(t)}{a^2(t)}dt^2+a^2(t)\delta_{ij}dx^idx^j.\end{equation}Here
$N(t)$ is the lapse function and $a(t)$ represents the scale factor
of the universe. The square of the scale factor dividing the lapse
function turns out to simplify the calculations and makes the
Hamiltonian quadratic. Now, it is easy to show that the effective
Lagrangian of the model can be written in the form
\begin{equation}\label{D}
{\cal L}=\frac{1}{N}\left(-\frac{1}{2} a^2
\dot{a}^2+\frac{1}{2}a^4
\dot{\phi}^2\right)-Na^2V(\phi).\end{equation}To simplify the
above Lagrangian, let us introduce a new set of variables
\cite{32}
\begin{equation}\label{E}
x=\frac{a^2}{2}\cosh \alpha \phi,\hspace{.5cm}y=\frac{a^2}{2}\sinh
\alpha \phi,\end{equation}where $\alpha$ is a positive constant. In
terms of these new variables the Lagrangian (\ref{D}) takes the form
\begin{equation}\label{F}
{\cal
L}=\frac{1}{2N}\left(\dot{y}^2-\dot{x}^2\right)-2N\left(x-y\right)e^{\alpha
\phi}V(\phi).\end{equation}From now on, we choose an exponential
potential
\begin{equation}\label{G}
V(\phi)=\frac{V_0}{2}e^{-\alpha \phi},\end{equation}which simplifies
the last term in the Lagrangian (\ref{F}) leading to
\begin{equation}\label{H}
{\cal
L}=\frac{1}{2N}\left(\dot{y}^2-\dot{x}^2\right)-NV_0\left(x-y\right),\end{equation}with
the corresponding Hamiltonian constraint written as
\begin{equation}\label{I}
{\cal
H}=-\frac{1}{2}p_x^2+\frac{1}{2}p_y^2+V_0\left(x-y\right).\end{equation}Note
that the minisuperspace of the above model is a two-dimensional
manifold  $0<a<\infty$, $-\infty<\phi<+\infty$. According to
\cite{33}, its nonsingular boundary is the line $a=0$ with
$|\phi|<+\infty$, while at the singular boundary, at least one of
the two variables is infinite. In terms of the variables $x$ and
$y$, introduced in (\ref{E}), the minisuperspace is recovered by
$x>0$, $x>|y|$, and the nonsingular boundary may be represented by
$x=y=0$.
\section{Classical cosmology}
The classical and quantum solutions of the model described by
Hamiltonian (\ref{I}) can be easily obtained. Since our aim here is
to compare the commutative solutions with noncommutative and GUP
solutions, in what follows we consider commutative, noncommutative
and GUP classical cosmologies, and compare the results with each
other. In the next section we shall deal with the quantum cosmology
of the model.
\subsection{Commutative case}
The Poisson brackets for the classical phase space variables are
\begin{equation}\label{J}
\left\{x_i,x_j\right\}=\left\{p_i,p_j\right\}=0,\hspace{.5cm}\left\{x_i,p_j\right\}=\delta_{ij},\end{equation}where
$x_i(i=1,2)=x,y$ and $p_i(i=1,2)=p_x, p_y$. Therefore, the
equations of motion become (in $N=1$ gauge)
\begin{equation}\label{K}
\dot{x}=\left\{x,{\cal
H}\right\}=-p_x,\hspace{.5cm}\dot{p_x}=\left\{p_x,{\cal
H}\right\}=-V_0,\end{equation}
\begin{equation}\label{L}
\dot{y}=\left\{y,{\cal
H}\right\}=p_y,\hspace{.5cm}\dot{p_y}=\left\{p_y,{\cal
H}\right\}=V_0,\end{equation}Equations (\ref{K}) and (\ref{L}) can
be immediately integrated to yield
\begin{equation}\label{M}
x(t)=\frac{1}{2}V_0
t^2-p_{0x}t+x_0,\hspace{.5cm}p_x(t)=-V_0t+p_{0x},\end{equation}
\begin{equation}\label{N}
y(t)=\frac{1}{2}V_0
t^2+p_{0y}t+y_0,\hspace{.5cm}p_y(t)=V_0t+p_{0y}.\end{equation}Now,
these solutions must satisfy the zero energy condition, ${\cal
H}=0$. Thus, substitution of equations (\ref{M}) and (\ref{N}) into
(\ref{I}) gives a relation between integration constants as
\begin{equation}\label{O}
p_{0y}^2-p_{0x}^2=2V_0 (y_0-x_0).\end{equation}Equations (\ref{M})
and (\ref{N}) are like the equation of motion for a particle moving
in a plane with its acceleration components equal to $V_0$, while
$-p_x(t)$ and $p_y(t)$ play the role of its velocity. Note that the
condition $x>0$ implies that $p_{0x}^2-2V_0x_0<0$, thus, equation
(\ref{O}) results in $p_{0y}^2-2V_0y_0<0$, which means that $y>0$.
Therefore, in classical cosmology only half of the minisuperspace:
$x>y>0$ or $(a>0, \phi >0)$ is recovered by the dynamical variables
$x(t)$ and $y(t)$. Now, using relations (\ref{E}) we can find the
scale factor and dilaton field as (to get a more simple form we take
$x_0=y_0$ and $p_{0x}=p_{0y}$ which of course satisfy the condition
(\ref{O}))
\begin{equation}\label{O1}
a(t)=\left[8|p_{0x}|V_0t^3+16x_0|p_{0x}|t\right]^{1/4},\end{equation}
\begin{equation}\label{O2}
\phi
(t)=\frac{1}{2\alpha}\ln\left(\frac{V_0t^2+2x_0}{2|p_{0x}|}\right).\end{equation}
The limiting behavior of $a(t)$ and $\phi (t)$ in the early and late
times is then as follows
\begin{equation}\label{O3}
a(t)\sim t^{1/4},\hspace{.5cm}\phi(t)\sim
\mbox{const.},\hspace{.5cm}t<<1,\end{equation}
\begin{equation}\label{O4}
a(t)\sim t^{3/4},\hspace{.5cm}\phi(t)\sim \ln
t,\hspace{.5cm}t>>1.\end{equation}A remark about the above analyze
is that we use a nonstandard parametrization of FRW metric, this is
done in order to simplify the calculations and have manageable
Lagrangian for the noncommutative deformation. As is well-known
usually the introduction of the lapse function gives a new
parametrization of time, but if $N(t)=1$ one returns to the usual
cosmic time where in our parametrization this is not the case.
Therefore, let us translate these results in terms of the cosmic
time $\tau$. Using its relationship with our time parameter $t$,
that is
\begin{equation}\label{O5}
d \tau=\frac{1}{a(t)}dt,\end{equation}we obtain
\begin{equation}\label{O6}
\tau \sim
t^{3/4}\hspace{.5cm}t<<1,\hspace{.5cm}\mbox{and}\hspace{.5cm}\tau
\sim t^{1/4}\hspace{.5cm}t>>1.\end{equation}Therefore, the behavior
of scale factor and the dilatonic field in the early and late
(cosmic) times is as
\begin{equation}\label{O7}
a(\tau)\sim \tau^{1/3},\hspace{.5cm}\phi(\tau) \sim
\mbox{const.}\hspace{.5cm}\tau <<1,\end{equation}
\begin{equation}\label{O8}
a(\tau)\sim \tau^{3},\hspace{.5cm}\phi(\tau) \sim \ln \tau
\hspace{.5cm}\tau >>1.\end{equation}We see that in the usual
commutative phase space of our model the scale factor has a
decelerated expansion in early times while undergoes an
accelerated phase in its late time evolution due to a constant and
growing with time dilatonic field respectively. These results are
comparable with those that are presented in the last paper of
\cite{25} where in which the authors used the gauge $d\tau=a^3dt$.
\subsection{Noncommutative case}
Let us now concentrate on the noncommutativity concepts in
classical cosmology. Noncommutativity in classical physics
\cite{23} is described by a deformed product, also known as the
Moyal product law between two arbitrary functions of position and
momenta as
\begin{equation}\label{Q}
(f*_{\alpha}g)(x)=\exp\left[\frac{1}{2}\alpha^{ab}\partial^{(1)}_a
\partial^{(2)}_b\right]f(x_1)g(x_2)|_{x_1=x_2=x},\end{equation}such
that
\begin{equation}\label{R}
\alpha_{ab}=\left(%
\begin{array}{cc}
\theta_{ij} & \delta_{ij}+\sigma_{ij} \\
-\delta_{ij}-\sigma_{ij} & \beta_{ij} \\
\end{array}%
\right),\end{equation}where the $N\times N$ matrices $\theta$ and
$\beta$ are assumed to be antisymmetric with $2N$ being the
dimension of the classical phase space, represents the
noncommutativity in coordinates and momenta, respectively. With this
product law, the deformed Poisson brackets can be written as
\begin{equation}\label{S}
\{f,g\}_\alpha=f*_\alpha g-g*_\alpha f.\end{equation}A simple
calculation shows that
\begin{equation}\label{T}
\{x_i,x_j\}_\alpha=\theta_{ij},\hspace{.5cm}\{x_i,p_j\}_\alpha=\delta_{ij}+\sigma_{ij},\hspace{.5cm}\{p_i,p_j\}_\alpha=\beta_{ij}.
\end{equation}Now, consider the following transformations on the classical phase-space
\begin{equation}\label{U}
x'_i=x_i-\frac{1}{2}\theta_{ij}p^j,\hspace{.5cm}p'_i=p_i+\frac{1}{2}\beta_{ij}x^j.
\end{equation}It can easily be checked that if $(x_i, p_j )$ obey the usual Poisson algebra (\ref{J}), then
\begin{equation}\label{V}
\{x'_i,x'_j\}=\theta_{ij},\hspace{.5cm}\{x'_i,p'_j\}=\delta_{ij}+\sigma_{ij},\hspace{.5cm}\{p'_i,p'_j\}=\beta_{ij},
\end{equation}where $\sigma_{ij}=-\frac{1}{8}\left(\theta_i^k\beta_{kj}+\beta_i^k\theta_{kj}\right)$.
These commutative relations are the same as (\ref{T}).
Consequently, for introducing noncommutativity, it is more
convenient to work with Poisson brackets (\ref{V}) than
$\alpha$-star deformed Poisson brackets (\ref{T}). It is important
to note that the relations represented by equations (\ref{T}) are
defined in the spirit of the Moyal product given above. However,
in the relations defined by (\ref{V}), the variables $(x_i, p_j )$
obey the usual Poisson bracket relations so that the two sets of
deformed and ordinary Poisson brackets represented by relations
(\ref{T}) and (\ref{V}) should be considered as distinct.

In this work we consider a noncommutative phase space in which
$\beta_{ij}= 0$ and so that $\sigma_{ij} = 0$, i.e. the Poisson
brackets of the phase-space variables are as follows
\begin{equation}\label{W}
\left\{x_{nc},y_{nc}\right\}=\theta,\hspace{.5cm}\left\{x_{inc},p_{jnc}\right\}=\delta_{ij},\hspace{.5cm}\left\{p_{inc},p_{jnc}\right\}=0.
\end{equation}With the noncommutative phase space defined above, we consider the Hamiltonian of the
noncommutative model as having the same functional form as
equation (\ref{I}), but in which the dynamical variables satisfy
the above-deformed Poisson brackets, that is
\begin{equation}\label{X}
{\cal
H}_{nc}=-\frac{1}{2}p_{xnc}^2+\frac{1}{2}p_{ync}^2+V_0\left(x_{nc}-y_{nc}\right).\end{equation}Therefore,
the equations of motion read
\begin{equation}\label{Y}
\dot{x_{nc}}=\left\{x_{nc},{\cal H}_{nc}\right\}=-p_{xnc}-\theta
V_0,\hspace{.5cm}\dot{p_{xnc}}=\left\{p_{xnc},{\cal
H}_{nc}\right\}=-V_0,\end{equation}
\begin{equation}\label{Z}
\dot{y_{nc}}=\left\{y_{nc},{\cal H}_{nc}\right\}=p_{ync}-\theta
V_0,\hspace{.5cm}\dot{p_{ync}}=\left\{p_{ync},{\cal
H}_{nc}\right\}=V_0.\end{equation}The above equations are similar
to equations (\ref{K}) and (\ref{L}) in the commutative case.
Their solutions are therefore as follows
\begin{equation}\label{AB}
x_{nc}=\frac{1}{2}V_0t^2-\left(p_{0x}+\theta
V_0\right)t+x_0,\hspace{.5cm}p_{xnc}=-V_0t+p_{0x},\end{equation}
\begin{equation}\label{AC}
y_{nc}=\frac{1}{2}V_0t^2+\left(p_{0y}-\theta
V_0\right)t+y_0,\hspace{.5cm}p_{ync}=V_0t+p_{0y}.\end{equation}The
requirement that these solutions must satisfy the noncommutative
Hamiltonian constraint ${\cal H}_{nc}=0$, gives us again the
relation (\ref{O}) between integration constants. As mentioned
before, instead of dealing with the noncommutative variables we can
construct, with the help of transformations (\ref{U}), a set of
commutative dynamical variables $x,y$ obeying the usual Poisson
brackets (\ref{J}) which, for the problem at hand read
\begin{eqnarray}\label{AD}
p_{x_{nc}}&=&p_x,\hspace{.5cm}p_{y_{nc}}=p_y,\nonumber\\
x_{nc}&=&x-\frac{1}{2}\theta
p_y,\hspace{.5cm}y_{nc}=y+\frac{1}{2}\theta p_x.
\end{eqnarray}In terms of these commutative variables the Hamiltonian takes the form
\begin{equation}\label{AE}
{\cal
H}=-\frac{1}{2}p_x^2+\frac{1}{2}p_y^2+V_0\left(x-y\right)-\frac{1}{2}\theta
V_0\left(p_x+p_y\right).\end{equation}Therefore, we have the
following equations of motion
\begin{equation}\label{AF}
\dot{x}=\left\{x,{\cal H}\right\}=-p_x-\frac{1}{2}\theta
V_0,\hspace{.5cm}\dot{p_x}=\left\{p_x,{\cal
H}\right\}=-V_0,\end{equation}
\begin{equation}\label{AG}
\dot{y}=\left\{y,{\cal H}\right\}=p_y-\frac{1}{2}\theta
V_0,\hspace{.5cm}\dot{p_y}=\left\{p_y,{\cal
H}\right\}=V_0.\end{equation}The solutions of the above equations
can be straightforwardly obtained in the same manner as that of
system (\ref{K})-(\ref{L}). It is easy to check that the action of
transformations (\ref{AD}) on the solutions of system
(\ref{AF})-(\ref{AG}) is to recover solutions (\ref{AB})-(\ref{AC}).
We see that the effects of noncommutative parameter $\theta$ appears
only in the initial velocity of the evolution. This means that
noncommutativity in phase space shows itself in the early epoch of
the cosmic evolution and when time grows the differences between
commutative solutions (\ref{M}), (\ref{N}) and noncommutative
solutions (\ref{AB}), (\ref{AC}) disappear. To make this issue more
clear, let us return to the variables $a(t)$ and $\phi(t)$ using the
transformation (\ref{E}). Choosing again $x_0=y_0$ and
$p_{0x}=p_{0y}$ we obtain
\begin{equation}\label{AG1}
a_{nc}(t)=\left[8|p_{0x}|V_0t^3+16\theta
|p_{0x}|t^2+16x_0|p_{0x}|t\right]^{1/4},\end{equation}
\begin{equation}\label{AG2}
\phi_{nc}(t)=\frac{1}{2\alpha}\ln \left|\frac{V_0t^2-2\theta
V_0t+2x_0}{2p_{0x}}\right|.\end{equation}The late time ($t>>1$)
behavior of $a_{nc}(t)$ and $\phi_{nc}(t)$ is the same as
(\ref{O4}). On the other hand in the regime $t<<1$, considering
$\theta$-term in (\ref{AG1}) and (\ref{AG2}) we obtain
\begin{equation}\label{AG3}
a_{nc}(t)\sim \theta^{1/4}t^{1/2},\hspace{.5cm}\phi_{nc}(t)\sim
\ln(\theta t),\hspace{.5cm}t<<1.\end{equation}In this limit the
cosmic time $d\tau=\frac{1}{a}dt$ takes the form
\begin{equation}\label{AG4}
\tau \sim \theta^{-1/4}t^{1/2},\end{equation}and then the early
(cosmic) time behavior of the scale factor and the dilatonic field
is as follows
\begin{equation}\label{AG5}
a_{nc}(\tau) \sim \theta^{1/2}\tau,\hspace{.5cm}\phi_{nc}(\tau) \sim
\ln (\theta^{3/2}\tau^2),\hspace{.5cm}\tau<<1.\end{equation}We see
that noncommutativity causes a uniform expansion (not decelerated
expansion) in the early times of cosmic evolution.
\subsection{Classical cosmology with GUP}
In more than one dimension a natural generalization of equation
(\ref{BB}) is defined by the following commutation relations
\cite{9}
\begin{equation}\label{AH}
\left[x_i,p_j\right]=i\left(\delta_{ij}+\beta
\delta_{ij}p^2+\beta'p_ip_j\right),\end{equation}where $p^2=\sum
p_ip_i$ and $\beta,\beta'>0$ are considered as small quantities of
first order. Also, assuming that
\begin{equation}\label{AI}
\left[p_i,p_j\right]=0,\end{equation} the commutation relations
for the coordinates are obtained as
\begin{equation}\label{AJ}
\left[x_i,x_j\right]=i\frac{(2\beta-\beta')+(2\beta+\beta')\beta
p^2}{1+\beta p^2}\left(p_ix_j-p_jx_i\right).\end{equation}As it is
clear from the above expression, the coordinates do not commute.
This means that to construct the Hilbert space representations,
one cannot work in position space. It is therefore more convenient
to work in momentum space. However, since in quantum cosmology the
wave function of the universe in momentum space has no suitable
interpretation, we restrict ourselves to the special case
$\beta'=2\beta$. As one can see immediately from equation
(\ref{AJ}), the coordinates commute to first order in $\beta$ and
thus a coordinate representation can be defined. Now, it is easy
to show that the following representation of the momentum operator
in position space satisfies relations (\ref{AH}) and (\ref{AI})
(with $\beta'=2\beta$) to first order in $\beta$
\begin{equation}\label{AL}
p_i=-i\left(1-\frac{\beta}{3}\frac{\partial^2}{\partial
x_i^2}\right)\frac{\partial}{\partial x_i}.\end{equation}A comment
on the above issue is that applying the GUP to a curved background
such as a cosmological model needs some modifications \cite{34}.
Here, since we apply the GUP to the minisuperspace variables $x,
y$ which correspond to a Minkowskian metric, we can safely use the
above expressions without any modifications. Now, it is possible
to realize equations (\ref{AH})-(\ref{AL}) from the following
commutation relations between position and momentum operators
\begin{equation}\label{AM}
\left[x,p_x\right]=i\left(1+\beta p^2+2\beta
p_x^2\right),\hspace{.5cm}\left[y,p_y\right]=i\left(1+\beta
p^2+2\beta p_y^2\right),\end{equation}
\begin{equation}\label{AN}
\left[x,p_y\right]=\left[y,p_y\right]=2i\beta p_x
p_y,\end{equation}
\begin{equation}\label{AO}
\left[x_i,x_j\right]=\left[p_i,p_j\right]=0,\hspace{.5cm}x_i(i=1,2)=x,y,\hspace{.5cm}p_i(i=1,2)=p_x,p_y.\end{equation}
Now, before quantizing the model in the GUP framework in the next
section, we would like to investigate the effects of classical
version of GUP, i.e. classical version of commutation relations
(\ref{AM})-(\ref{AO}) on the above cosmology. As is well known, in
classical limit the quantum mechanical commutators should be
replaced by the classical Poisson brackets as
$\left[P,Q\right]\rightarrow i \hbar \left\{P,Q\right\}$. Thus, the
GUP in classical phase space changes the Poisson algebra (\ref{J})
into their deformed forms as \footnote{Such deformed Poisson algebra
is used in \cite{Ben} to investigate effects of the deformation on
the classical orbits of particles in a central force field and on
the Kepler third law. Also, the stability of planetary circular
orbits in the framework of such deformed Poisson brackets is
considered in \cite{Ku}. Note that here we deal with modifications
of a classical cosmology that become important only at the Planck
scale, where the classical description is no longer appropriate and
a quantum model is required. However, before quantizing the model we
shall provide a deformed classical cosmology. In this classical
description of the universe in transition from commutation relation
(\ref{BB}) to its Poisson bracket counterpart we keep the parameter
$\beta$ fix as $\hbar \rightarrow 0$. In string theory this means
that the string momentum scale is fixed when its length scale
approaches the zero.}
\begin{equation}\label{AP}
\left\{x,p_x\right\}=1+\beta p^2+2\beta
p_x^2,\hspace{.5cm}\left\{y,p_y\right\}=1+\beta p^2+2\beta
p_y^2,\end{equation}
\begin{equation}\label{AQ}
\left\{x,p_y\right\}=\left\{y,p_x\right\}=2\beta p_x
p_y,\end{equation}
\begin{equation}\label{AR}
\left\{x_i,x_j\right\}=\left\{p_i,p_j\right\}=0,\hspace{.5cm}x_i(i=1,2)=x,y,\hspace{.5cm}p_i(i=1,2)=p_x,p_y.\end{equation}
Therefore, the equations of motion read
\begin{equation}\label{AS}
\dot{x}=\left\{x,{\cal H}\right\}=-p_x\left(1-\beta
p^2\right),\hspace{.5cm}\dot{p_x}=\left\{p_x,{\cal
H}\right\}=-V_0\left[1+\beta\left(p_y-p_x\right)^2\right],\end{equation}
\begin{equation}\label{AT}
\dot{y}=\left\{y,{\cal H}\right\}=p_y\left(1+3\beta
p^2\right),\hspace{.5cm}\dot{p_y}=\left\{p_y,{\cal
H}\right\}=V_0\left[1+\beta\left(p_y-p_x\right)\left(3p_y+p_x\right)\right].\end{equation}
We see that the deformed classical cosmology form a system of
nonlinear coupled differential equations, which are not easy to
solve. Thus, to simplify it, we may make some approximations. From
equations (\ref{AS}) and (\ref{AT}), we get
\begin{equation}\label{AU}
\dot{p_x}+\dot{p_y}=2\beta
V_0\left(p_y^2-p_x^2\right),\end{equation}if in the first
approximation we neglect the right hand side of the above
equation, we obtain
\begin{equation}\label{AV}
\dot{p_x}+\dot{p_y}=0\Rightarrow
p_x+p_y=p_0=\mbox{Const}.\end{equation}Substituting this result in
equations (\ref{AS}) and (\ref{AT}), we are led to the following
decoupled equations for $p_x$ and $p_y$
\begin{equation}\label{AW}
\dot{p_x}=-V_0\left[1+\beta
\left(p_0-2p_x\right)^2\right],\end{equation}
\begin{equation}\label{AX}
\dot{p_y}=V_0\left[1+\beta
\left(4p_y^2-p_0^2\right)\right],\end{equation}where are immediately
integrable with the result
\begin{equation}\label{AY}
p_x(t)=\frac{1}{2}p_0-\frac{1}{2\sqrt{\beta}}\tan
2\sqrt{\beta}V_0\left(t+t_0\right),\end{equation}
\begin{equation}\label{AZ}
p_y(t)=\frac{1}{2}p_0+\frac{1}{2\sqrt{\beta}}\tan
2\sqrt{\beta}V_0\left(t+t_0\right).\end{equation}  Substituting
these results into the first equations of the system (\ref{AS})
and (\ref{AT}), we can obtain $x(t)$ and $y(t)$ as
\begin{eqnarray}\label{BA}
x(t)&=&\frac{p_0}{4}(p_0^2\beta
-3)t+\frac{p_0}{8V_0\sqrt{\beta}}\tan 2V_0
\sqrt{\beta}(t+t_0)+\left(\frac{p_0^2}{8V_0}-\frac{3}{8V_0
\beta}\right)\ln \left[ \cos 2V_0
\sqrt{\beta}(t+t_0)\right]\nonumber
\\ &&-\frac{1}{16\beta V_0}\tan ^2 2V_0
\sqrt{\beta}(t+t_0),\end{eqnarray}
\begin{eqnarray}\label{BC}
y(t)&=&\frac{p_0}{4}(3p_0^2\beta
-1)t+\frac{3p_0}{8V_0\sqrt{\beta}}\tan 2V_0
\sqrt{\beta}(t+t_0)-\left(\frac{3p_0^2}{8V_0}-\frac{1}{8V_0
\beta}\right)\ln \left[ \cos 2V_0
\sqrt{\beta}(t+t_0)\right]\nonumber
\\ &&+\frac{3}{16\beta V_0}\tan ^2 2V_0
\sqrt{\beta}(t+t_0).\end{eqnarray} It is easy to see that in the
limit $\beta \rightarrow 0$, with a suitable choice of $t_0$ in
terms of $p_{0x}$, $p_{0y}$ and $V_0$, we can recover the ordinary
classical cosmology (\ref{M}) and (\ref{N}). A comment on the above
solutions is that the effects of GUP are important not only in the
early but also at late times of the cosmic evolution. In fact, these
solutions show that in the GUP framework the quantum gravitational
effects may be detected also in large scales.
\section{Quantization of the model}
Now, let us quantize the model described above. As in the
classical cosmology, here for comparison purposes between ordinary
commutative, noncommutative and GUP, we study the quantum
cosmology of the model in these frameworks separately and compare
the results.
\subsection{Commutative quantum cosmology}
We first discuss the commutative quantum cosmology of our model. For
this purpose we quantize the dynamical variables of the model with
the use of canonical quantization procedure that leads to the
Wheeler- DeWitt (WD) equation, ${\cal H}\Psi=0$. Here, ${\cal H}$ is
the operator form of the Hamiltonian given by (\ref{I}), and $\Psi$
is the wave function of the universe, a function of spatial geometry
and matter fields, if they exist. With replacement $p_x\rightarrow
-i \partial/\partial x$ and similarly for $p_y$ in (\ref{I}), the WD
equation reads
\begin{equation}\label{BD}
\left[\frac{\partial^2}{\partial x^2}-\frac{\partial^2}{\partial
y^2}+2V_0 \left(x-y\right)\right]\Psi(x,y)=0.\end{equation} The
solutions of the above differential equation are separable and may
be written in the form $\Psi(x,y)=X(x)Y(y)$, leading to
\begin{equation}\label{BE}
\frac{d^2X}{dx^2}+\left(2V_0x-\nu\right)X=0,\hspace{.5cm}\frac{d^2Y}{dy^2}+\left(2V_0y-\nu\right)Y=0,\end{equation}
where $\nu$ is a separation constant. Equations (\ref{BE}) have
well- known solutions in terms of Airy functions $\mbox{Ai}(z)$
and $\mbox{Bi}(z)$. The functions $\mbox{Bi}(z)$ are usually
omitted because of their divergent behavior in the limit
$z\rightarrow \infty$. Therefore, the eigenfunctions of the WD
equation can be written as
\begin{equation}\label{BF}
\Psi_{\nu}(x,y)=\mbox{Ai}\left(\frac{\nu
-2V_0x}{(2V_0)^{2/3}}\right)\mbox{Ai}\left(\frac{\nu-
2V_0y}{(2V_0)^{2/3}}\right).\end{equation} Now, we impose the
boundary condition on these solutions such that at the nonsingular
boundary (at $a=0$ and $|\phi|<\infty$) the wave function vanishes
\cite{33}
\begin{equation}\label{BG}
\Psi(a=0,\phi)=0\Rightarrow \Psi(x=0,y=0)=0,\end{equation}which
yields
\begin{equation}\label{BH}
\mbox{Ai}\left(\frac{\nu}{(2V_0)^{2/3}}\right)=0\Rightarrow
\nu_n=(2V_0)^{2/3}\alpha_n,\end{equation} where $\alpha_n$ is the
$n$th zero of the Airy function $\mbox{Ai}(z)$. We may now write
the general solution of the WD equation as a superposition of its
eigenfunctions
\begin{equation}\label {BI}
\Psi(x,y)=\sum_{n=1}^{\infty}c_n\mbox{Ai}\left(\frac{\nu_n
-2V_0x}{(2V_0)^{2/3}}\right)\mbox{Ai}\left(\frac{\nu_n-
2V_0y}{(2V_0)^{2/3}}\right).\end{equation}Figure 1 shows the square
of wave function of the commutative quantum universe. As is clear
from this figure the wave function peaks symmetrically around $y=0$.
The largest peaks correspond to some nonzero values $x_0$ for $x$
and $\pm y_0$ for $y$. This means that there are different possible
states (correspond to positive and negative dilaton) from which our
present universe could have evolved and tunneled in the past, from
one state to another.
\begin{figure}\begin{center}
\epsfig{figure=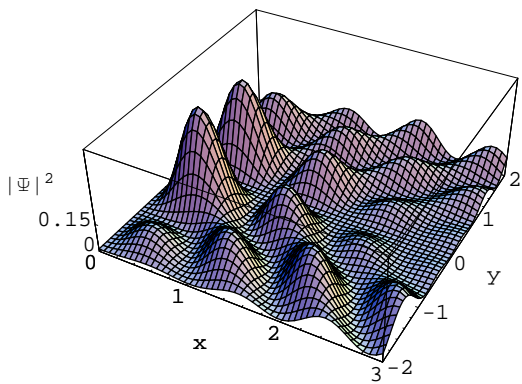,width=7cm}\caption{\footnotesize The
square of wave function in the commutative case. We take the
numerical value $V_0=1$.} \label{fig1}\end{center}
\end{figure}
\subsection{Noncommutative quantum cosmology}
To study noncommutativity at the quantum level, we follow the same
procedure as before, namely the canonical transition from
classical to quantum mechanics by replacing the Poisson brackets
with the corresponding Dirac commutators
$\left\{,\right\}\rightarrow -i\left[,\right]$. Thus, the
commutation relations between our dynamical variables should be
modified as follows
\begin{equation}\label{BJ}
\left[x_{nc},y_{nc}\right]=i\theta,\hspace{.5cm}\left[x_{nc},p_x\right]=\left[y_{nc},p_y\right]=i.\end{equation}The
corresponding WD equation can be obtained by modification of the
operator product in (\ref{BD}) with the Moyal deformed product
\cite{25}
\begin{equation}\label{BK}
\left[-\frac{1}{2}p_x^2+\frac{1}{2}p_y^2+V_0\left(x-y\right)\right]\ast
\Psi(x,y)=0,\end{equation}Using the definition of the Moyal
product (\ref{Q}), it may be shown that
\begin{equation}\label{BL}
f(x,y)\ast \Psi(x,y)=f(x_{nc},y_{nc})\Psi(x,y),\end{equation}where
the relations between the noncommutative variables $x_{nc}$,
$y_{nc}$ and commutative variables $x$, $y$ are given by (\ref{AD}).
Therefor, the noncommutative version of the WD equation can be
written as
\begin{equation}\label{BM}
\left[\frac{\partial^2}{\partial x^2}-\frac{\partial^2}{\partial
y^2}+i\theta V_0\left(\frac{\partial}{\partial
x}+\frac{\partial}{\partial
y}\right)+2V_0\left(x-y\right)\right]\Psi(x,y)=0.\end{equation}We
again separate the solutions into the form $\Psi(x,y)=X(x)Y(y)$,
which leads to the following equations for the functions $X(x)$
and $Y(y)$ with a separation constant $\nu$
\begin{eqnarray}\label{BN}
\frac{d^2X}{dx^2}+i\theta V_0\frac{dX}{dx}+\left(2V_0 x-\nu\right)X&=&0,\nonumber\\
\frac{d^2Y}{dy^2}-i\theta V_0\frac{dY}{dy}+\left(2V_0
y-\nu\right)Y&=&0.
\end{eqnarray}The solutions of equations (\ref{BN}) can be written
in terms of Airy functions as
\begin{eqnarray}\label{BO}
X(x)&=&e^{-\frac{i}{2} V_0 \theta
x}\mbox{Ai}\left(\frac{\nu-\frac{1}{4}V_0^2\theta
^2-2V_0x}{(2V_0)^{2/3}}\right),\nonumber\\Y(y)&=&e^{\frac{i}{2}
V_0\theta y}\mbox{Ai}\left(\frac{\nu-\frac{1}{4}V_0^2\theta
^2-2V_0y}{(2V_0)^{2/3}}\right),\end{eqnarray}where to recover the
commutative solutions in the case of $\theta=0$, we have omitted
the functions $\mbox{Bi}(z)$. Thus the eigenfunctions of
noncommutative WD equation are as follows
\begin{equation}\label{BP}
\Psi_{\nu}(x,y)=e^{\frac{i}{2}V_0\theta
(y-x)}\mbox{Ai}\left(\frac{\nu-\frac{1}{4}V_0^2\theta
^2-2V_0x}{(2V_0)^{2/3}}\right)\mbox{Ai}\left(\frac{\nu-\frac{1}{4}V_0^2\theta
^2-2V_0y}{(2V_0)^{2/3}}\right).\end{equation}Note that in the
context of our noncommutative model choosing the boundary
condition (\ref{BG}) is not trivial and instead we construct the
general solution of WD equation as a superposition of
eigenfunctions in the form
\begin{equation}\label{BQ}
\Psi(x,y)=\int_{-\infty}^{+\infty}C(\nu)\Psi_{\nu}(x,y)d
\nu,\end{equation}where $C(\nu)$ can be chosen as a shifted Gaussian
weight function $e^{-a(\nu -b)^2}$, see Refs. \cite{25}. Figure 2
shows the square of the wave function in the noncommutative case. We
see that in this case the peaks occur only for positive values of
$y$ (or positive values of dilaton). Also, noncommutivity causes a
shift in the minimum of the values of $x$ corresponding to the
spacial volume.
\begin{figure}\begin{center}
\epsfig{figure=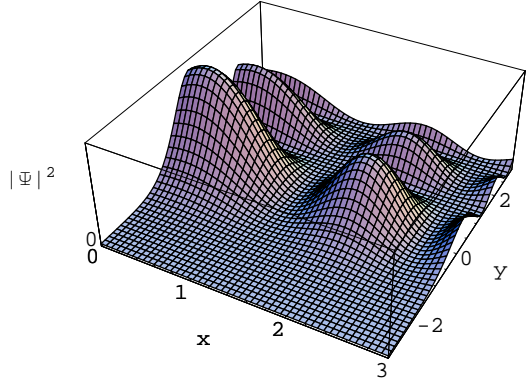,width=7cm}\caption{\footnotesize The
square of wave function in the noncommutative case. We take the
numerical values $V_0=1$ and $\theta =2$.}
\label{fig2}\end{center}
\end{figure}
\subsection{Quantum cosmology in GUP framework}
In this subsection we focus attention in the study of the quantum
cosmology of our model based on the GUP formalism reviewed in the
previous section. The corresponding commutation relations are
given by (\ref{AM})-(\ref{AO}). As we have mentioned in the
previous section, in the special case when $\beta'=2\beta$, we
have the following representations for $p_x$ and $p_y$ in the
$x-y$ space which fulfill the commutation relations
(\ref{AM})-(\ref{AO})
\begin{equation}\label{BR}
p_x=-i\left(1-\frac{\beta}{3}\frac{\partial^2}{\partial
x^2}\right)\frac{\partial}{\partial
x},\hspace{.5cm}p_y=-i\left(1-\frac{\beta}{3}\frac{\partial^2}{\partial
y^2}\right)\frac{\partial}{\partial y}.\end{equation}Now, using
these representations for the momenta in Hamiltonian (\ref{I}), the
WD equation can be written, up to the first order in $\beta$ as
\begin{equation}\label{BS}
\left[-\frac{2}{3}\beta \frac{\partial^4}{\partial
x^4}+\frac{\partial^2}{\partial x^2}+\frac{2}{3}\beta
\frac{\partial^4}{\partial y^4}-\frac{\partial^2}{\partial
y^2}+2V_0\left(x-y\right)\right]\Psi(x,y)=0.
\end{equation}We again separate the solutions into the form
$\Psi(x,y)=X(x)Y(y)$, leading to
\begin{equation}\label{BT}
-\frac{2}{3}\beta
\frac{d^4Z_i}{dz_i^4}+\frac{d^2Z_i}{dz_i^2}+\left(2V_0z_i-\nu\right)Z_i=0,\hspace{.5cm}Z_i(i=1,2)=X,Y,
\hspace{.5cm}z_i(i=1,2)=x,y,\end{equation} where $\nu$ is the
separation constant as before. We cannot solve the above fourth
order equations analytically, but we can provide an approximation
method which in its domain of validity, we need to solve a second
order differential equation. Taking $\beta=0$ in equations
(\ref{BT}) yields the ordinary WD equation where their solutions are
given by (\ref{BF}). In the case when $\beta\neq 0$, note that the
effects of $\beta$ are important at the Planck scales, i. e. in
cosmology language in the very early universe, that is, when the
scale factor is small, which in our model means $x,y\sim 0$.  Thus,
if we use the solutions (\ref{BF}) in the $\beta$-term of
(\ref{BT}),we may obtain some approximate analytical solutions in
the region $x,y\rightarrow 0$. To this end, we write the limiting
behavior of the solutions (\ref{BF}) in the region $x,y \sim 0$ as
\begin{equation}\label{BU}
\mbox{Ai}\left(\frac{\nu-2V_0z}{(2V_0)^{2/3}}\right)\rightarrow
c_0+c_1z+c_2z^2+c_3z^3+c_4z^4+{\cal O}(z^5).\end{equation}Therefore,
we can replace the fourth derivative of $X(x)$ and $Y(y)$ in
equations (\ref{BT}) with a constant and thus are led to the
following equations
\begin{equation}\label{BV}
\frac{d^2Z_i}{dz_i^2}+\left(2V_0z_i-\nu\right)Z_i=\beta_0,\hspace{.5cm}Z_i(i=1,2)=X,Y,\hspace{.5cm}z_i(i=1,2)=x,y,
\end{equation}
in which $\beta_0=16c_4\beta$. The solutions of above equation can
be written in terms of Airy functions and Hypergeometric functions
$\hspace{.2cm}$$F_{q\hspace{-.6cm}p}\hspace{.4cm}\left(\left\{a_1,...,a_p\right\};\left\{b_1,...,b_q\right\};z\right)$
as
\begin{equation}\label{BW}
Z(z)=\mbox{Ai}\left(\frac{\nu-2V_0z}{(2V_0)^{2/3}}\right)+{\cal
A}V_0\nu \beta_0
\mbox{Ai}\left(\frac{\nu-2V_0z}{(2V_0)^{2/3}}\right)F_{2\hspace{-.5cm}1}\hspace{.4cm}\left(\frac{1}{3};
\frac{2}{3},\frac{4}{3};\frac{(\nu-2V_0z)^3}{
36V_0^2}\right)+...,
\end{equation}where ${\cal A}$ is
\[{\cal A}=\frac{32^{2/3}3^{5/6}\pi}{36\Gamma(2/3)},\] and ,..., denotes the terms that we have neglected in our
approximation proposal. We have also removed the Airy functions
$\mbox{Bi}(z)$ from the solutions to recover the solutions
(\ref{BF}) in the limit $\beta\rightarrow 0$. Thus, the
eigenfunctions of WD equation (\ref{BS}) read
\begin{eqnarray}\label{BX}
\Psi_{\nu}(x,y)&=&\left[\mbox{Ai}\left(\frac{\nu-2V_0x}{(2V_0)^{2/3}}\right)+{\cal
A}V_0\nu \beta_0
\mbox{Ai}\left(\frac{\nu-2V_0x}{(2V_0)^{2/3}}\right)F_{2\hspace{-.5cm}1}\hspace{.4cm}\left(\frac{1}{3};\frac{2}{3}
,\frac{4}{3};\frac{(\nu-2V_0x)^3}{
36V_0^2}\right)\right]\times
\nonumber\\&&\left[\mbox{Ai}\left(\frac{\nu-2V_0y}{(2V_0)^{2/3}}\right)+{\cal
A}V_0\nu \beta_0
\mbox{Ai}\left(\frac{\nu-2V_0y}{(2V_0)^{2/3}}\right)F_{2\hspace{-.5cm}1}\hspace{.4cm}\left(\frac{1}{3};\frac{2}{3}
,\frac{4}{3};\frac{(\nu-2V_0y)^3}{
36V_0^2}\right)\right].
\end{eqnarray}
Now, bearing in the mind that in our GUP framework, we have chosen
the GUP parameters $\beta'=2\beta$ such that the coordinates
commute, we can apply the boundary condition ({\ref{BG}), also on
the GUP wave function, which yields
\begin{equation}\label{BY}
\mbox{Ai}\left(\frac{\nu}{(2V_0)^{2/3}}\right)+{\cal A}V_0\nu
\beta_0
\mbox{Ai}\left(\frac{\nu}{(2V_0)^{2/3}}\right)F_{2\hspace{-.5cm}1}\hspace{.4cm}\left(\frac{1}{3};\frac{2}{3}
,\frac{4}{3};\frac{\nu^3}{
36V_0^2}\right)=0.\end{equation}Therefore, the general solution of
WD equation can be written as
\begin{eqnarray}\label{BZ}
\Psi(x,y)&=&\sum_n
c_n\left[\mbox{Ai}\left(\frac{\nu_n-2V_0x}{(2V_0)^{2/3}}\right)+{\cal
A}V_0\nu_n \beta_0
\mbox{Ai}\left(\frac{\nu_n-2V_0x}{(2V_0)^{2/3}}\right)F_{2\hspace{-.5cm}1}\hspace{.4cm}\left(\frac{1}{3}
;\frac{2}{3},\frac{4}{3};\frac{(\nu_n-2V_0x)^3}{
36V_0^2}\right)\right]\times
\nonumber\\&&\left[\mbox{Ai}\left(\frac{\nu_n-2V_0y}{(2V_0)^{2/3}}\right)+{\cal
A}V_0\nu_n \beta_0
\mbox{Ai}\left(\frac{\nu_n-2V_0y}{(2V_0)^{2/3}}\right)F_{2\hspace{-.5cm}1}\hspace{.4cm}\left(\frac{1}{3}
;\frac{2}{3},\frac{4}{3};\frac{(\nu_n-2V_0y)^3}{
36V_0^2}\right)\right],
\end{eqnarray}where $\nu_n$ are the zeros of equation
(\ref{BY}). In figure 3 we have plotted the square of wave function
when the phase space variables obey GUP relations, for small values
of $x$ and $y$. This figure shows only a possible state in the early
universe with a negative value for $y$ and a nonzero positive value
of $x$. Thus, in the context of GUP quantum cosmology our universe
emerges from a nonsingular state where the dilaton field has a
negative value.
\begin{figure}\begin{center}
\epsfig{figure=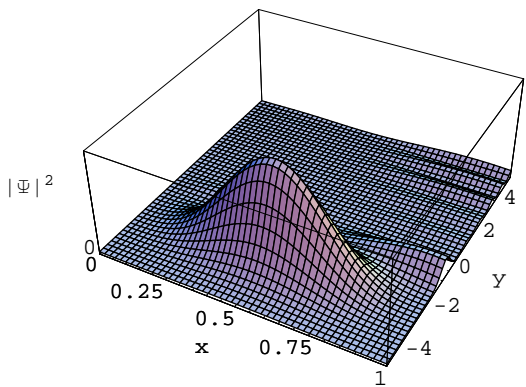,width=7cm}\caption{\footnotesize The
square of wave function in the GUP case. We take the numerical
values $V_0=1$ and $\beta_0=1$.} \label{fig3}\end{center}
\end{figure}
\section{Conclusions and comparison of the results}
In this paper we have studied the effects of noncommutativity and
generalized uncertainty relations in phase space, on classical and
quantum cosmology of a dilaton model with an exponential dilaton
potential. In the case of commutative phase space, the evolution of
the classical universe is like the motion of a particle (universe)
moving on a plane with a constant acceleration. We have shown that
in this case both dynamical variables $x$ and $y$ should be positive
which means that only half of the minisuperspace is recovered
through the evolution of the universe. In the case when quantum
cosmology is considered in the commutative phase space, we have seen
that the wave function of the universe peaks symmetrically around
$y=0$, which means that the present universe could have evolved from
different states with the same values for $x$ but different
symmetric values for $y$. In the case of noncommutative classical
cosmology, the solutions are like the commutative case with a little
difference, that is, the noncommutative parameter shows its effect
on the initial velocity of the evolution. On the other hand
noncommutative quantum cosmology predicts the emergence of the
universe from a positive value of $y$, that is, from a positive
value of the dilaton field. Finally, when the phase space variables
obey the GUP relations, the classical cosmology is described by
equations (\ref{BA}) and (\ref{BC}), which are more complicated than
commutative case. Also, we have presented approximate analytical
solutions of quantum cosmology in the GUP framework. These solutions
show only one possible state in the early universe with a negative
value for $y$ and a nonzero positive value of $x$. Thus, in the
context of GUP quantum cosmology the early universe emerges from a
nonsingular state where the dilaton field has a negative value.
\vspace{5mm}\newline \noindent {\bf
Acknowledgement}\vspace{2mm}\noindent\newline The author would like
to thank H. R. Sepangi for a careful reading of the manuscript and
useful comments.

\end{document}